\documentclass[journal,twocolumn]{IEEEtranTCOM}
\usepackage{graphicx,cite,epsfig}
\usepackage{amsmath} % assumes amsmath package installed
\usepackage{amssymb}  % assumes amsmath package installed
\DeclareMathOperator *{\argmin}{argmin}
\usepackage{pifont}
\usepackage{stmaryrd}
\usepackage{bbding}
\usepackage{subfigure}
\usepackage{algorithm}
\usepackage{algorithmic}
\usepackage{array}
\usepackage{amsthm}
\usepackage{url}
\usepackage{mathrsfs}
\usepackage{color, soul} %ÓÃcolor, ºÍ soul °ü
\usepackage{flushend}

\begin{document}

\pagestyle{empty}
\title{Grant-free Rateless Multiple Access: A Novel Massive Access Scheme for Internet of Things}

\author{
{\normalsize Zhaoyang Zhang, Xianbin Wang, Yu Zhang, and Yan Chen}
\thanks{This work was supported in part by National Key Basic Research Program of China (No. 2012CB316104), National Hi-Tech R\&D Program of China (No. 2014AA01A702), National Natural Science Foundation of China (No. 61371094, No. 61401391), the open project of Zhejiang Provincial Key Laboratory of Info. Proc., Commun. \& Netw., China, and Huawei Technologies Co., Ltd (YB2013120029).

Z. Zhang (Corresponding Author, {\tt email: ning\_ming@zju.edu.cn}) and X. Wang ({\tt email: wangxianbin@outlook.com}) are with College of Information Science and Electronic Engineering, Zhejiang University, China. Y. Zhang ({\tt email: zhangyu\_wing@hotmail.com}) is with College of Information Engineering, Zhejiang University of Technology, China and Zhejiang Provincial Key Laboratory of Information Proc., Commun. \& Netw., China. Y. Chen ({\tt email: bigbird.chenyan@huawei.com}) is with Huawei Technologies Co., Ltd, China.}

}

\maketitle
\thispagestyle{empty}
\vspace{-2cm}
\begin{abstract}
Rateless Multiple Access (RMA) is a novel non-orthogonal multiple access framework that is promising for massive access in Internet of Things (IoT) due to its high efficiency and low complexity. In the framework, after certain \emph{registration}, each active user respectively transmits to the access point (AP) randomly based on an assigned random access control function (RACf) until receiving an acknowledgement (ACK).
In this work, by exploiting the intrinsic access pattern of each user, we propose a grant-free RMA scheme, which no longer needs the registration process as in the original RMA, thus greatly reduces the signalling overhead and system latency. Furthermore, we propose a low-complexity joint iterative detection and decoding algorithm in which the channel estimation, active user detection, and information decoding are done simultaneously. Finally, we propose a method based on density evolution (DE) to evaluate the system performance.
\end{abstract}

\section{introduction}
Recently, massive access technology in Internet of Things (IoT) has been attracting more and more attention due to the great demand and big challenges confronted in system design and deployment \cite{M2M,Short}. In such a system, numerous always-on-line devices, probably 100 times more than those served by an up-to-date access point (AP)\cite{M2M}, require the system to support the massive connections. Furthermore, short packets (e.g., hundreds of bits or less) will constitute the majority of the traffic \cite{Short}, which necessitates the minimization of signalling overhead. Last but not least, in some applications such as Internet of Vehicles (IoV), the devices should be served with extremely low latency (e.g., 1ms). However, in the current Long Term Evolution (LTE) system, the uplink transmission is scheduled by the AP with a request-grant procedure, i.e., the user should send a scheduling request (SR) to the AP during the \emph{registration} procedure at first and then the AP performs scheduling to \emph{grant} resources to users in a centralized manner\cite{LTE}. If adopting this framework in IoT, the resultant signalling overhead and system latency will be totally unacceptable. These bring the necessity for novel \emph{grant-free} access mechanism in the physical (PHY) layer, based on which, the aforementioned request-grant procedure can be omitted and user identities and user data can be transmitted to the AP simultaneously. Such a protocol design is challenging since it needs to meet the above requirements at the same time.

Recently, a novel random massive access framework called rateless multiple access (RMA) was proposed (see \cite{RMA} and references therein). In this framework, after the \emph{registration}, instead of granting each active user with fixed resources elements (RE, e.g., certain subcarrier-time slot pair), the AP assigns to each user a random access control function (RACf) which enables them to share a block of REs in a random manner. In particular, for every RE, each user independently chooses a random number of coded symbols according to the assigned RACf, and then sends out their sum over it. At the AP, a low-complexity belief propagation (BP) algorithm is used to recover the original information. The capacity-approaching, highly flexible and low-complexity characteristics make the framework attractive for future large-scale networking. However, RMA still has its limitation when applied to IoT which in general consists of a large number of machine-type nodes, in which the \emph{registration} procedure will greatly reduce the efficiency and increase the system latency.

In this work, we propose a grant-free RMA scheme to overcome the above challenges. We exploit an intrinsic feature of RMA, i.e., each user has its own unique pseudo-random pattern for the access of REs, as a hidden clue to identify the user's activity within the current RE block. Since the number of active users is typically orders of magnitude smaller than the access pattern space, it results in certain \emph{sparsity} that can be exploited to do the active user detection. A low-complexity joint iterative detection and decoding algorithm based on BP is then proposed in which the channel estimation, the active user detection, and the information decoding are done simultaneously. Finally, we propose a method based on density evolution (DE) to evaluate the system performance.% which presents the relationship between the throughput and the worst-case decoding threshold of signal-to-noise ratio (SNR).

Note that the active user detection in sparse system has also been investigated in \cite{Zhu,Bockelmann1,Bockelmann3}. Compared with these works, the advantages of our scheme are two-fold: (i) Previous works require that the channel state information (CSI) are available to the AP while our scheme does not, which makes sense for short packet transmission in IoT since the acquisition of CSI often incurs large amount of signalling; (ii) Unlike the previous works that apply the relatively complicated maximum a \emph{posteriori} probability (MAP) detection, the proposed BP-based joint detection and decoding is of affordably low complexity, which makes it viable for massive access.

%In summary, the main contributions of this work are: \begin{enumerate}
%                                                       \item We propose the grant-free RMA scheme to address the challenge brought by short-packet and low-latency transmissions in IoT networks. Our scheme offers high efficiency with low complexity algorithm, thus is a viable candidate for massive access in the IoT system;
%                                                       \item We propose a novel DE analysis method to derive the relationship between the system throughput and the threshold SNR, which gives instructions to the practical system design. The proposed analysis method can be extended to other situations in which the influence of estimated CSI is analyzed.
%                                                     \end{enumerate}

%The rest of the paper is organized as follows. In Section \ref{scheme}, after summarizing the system model, we describe the grant-free access procedure and the joint active user detection and decoding algorithm. Section \ref{analysis} provides the performance analysis and Section \ref{numer} gives simulation results. At last, we conclude the paper in Section \ref{conclusion}. Table \uppercase\expandafter{\romannumeral1} lists the important variables used in this paper for easy reference.

\section{Protocol and Algorithm} \label{scheme}
Consider a cellular network in which massive potential users independently access the AP sporadically. All the users are perfectly synchronized according to the downlink beacon from the AP. When one user is active, it transmits to the AP a short data packet. Since the packet is relatively short, the links between users and the AP are assumed to keep static during each transmission cycle.
%\begin{table}
%\newcommand{\tabincell}[2]{\begin{tabular}{@{}#1@{}}#2\end{tabular}}
%  \centering
%    \caption{List of important variables}
%  \begin{tabular}{|c|c|}
%\hline
%\textbf{Variables} & \textbf{Descriptions}\\
%\hline
% $K$ & Number of potential users\\
% \hline
% $\mathcal{K}_a$ & Set of indices of active users\\
%  \hline
%  $T$ & Number of REs\\
% \hline
% $h_k$ & Channel gain from user $U_k$ to the AP \\
%\hline
% $p_\text{a}$ & User active probability\\
% \hline
% $(\bar{\mu}_{h_k},\bar{\xi}_{h_k})$ & Initial mean and variance information \\ &of User $k$'s channel gain\footnotemark[1]\\
%\hline
%$(\mu_x^n,\xi_x^n)$ & Estimated mean and variance of R.V. $x$ \\&at iteration $n$ \\
%\hline
%\end{tabular}
%\end{table}
%\footnotetext[1]{We assume that the AP knows some rough information about the mean and variance of channel gains so as to start the iterative detection process. In case they are unavailable, we can simply set them to be some proper non-zero values (e.g. $1$ and $10$, respectively) \cite{WO1}. The joint iterative algorithm can then re-estimate their exact values.
%}

\subsection{The Proposed Grant-free access protocol}
When the system is set up, user $k$, denoted by $U_k$, $k\in\{1,2,...K\}$, is assigned by the AP an RACf (please be referred to \cite{RMA} for the detailed design and optimization) as follows:
\begin{equation}\small		
\begin{aligned}
\label{degree_profile}
   \rho(x)=\sum_{d=0}^{d_\text{max}}p_{d}x^d
\end{aligned}
\end{equation}
where $\sum_{d=0}^{d_\text{max}} p_{d}=1$.
\begin{figure}[!ht]
\centering
\includegraphics[width=0.48\textwidth]{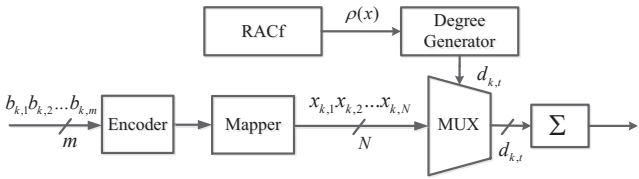}
\caption{Block diagram of the user side in the grant-free RMA.}% each active user pseudo-randomly transmits to the AP based on its RACf.}
\label{F_transmodel}
\end{figure}

At a given time, each user is active with probability $p_\text{a}$. If $U_k$ is inactive, it just keeps silent. Otherwise, as illustrated in Fig.\ref{F_transmodel}, it first encodes its message ($m$  bits) with a low-density parity-check (LDPC) encoder and maps the resultant coded bits to a symbol vector $\{x_{k,1}, x_{k,2}, ..., x_{k,N}\}$ with $\pm1$-valued indices. Then as the AP broadcasts a beacon to start the access period, for each RE $t ~(t\in\{1,2,...T\})$, $U_k$ first pseudo-randomly generates a degree $d_{k,t}$ from $0$ to $d_\text{max}$ with probability $p_{d_{k,t}}$ according to its RACf. If $d_{k,t} = 0$, it transmits nothing over that RE. Otherwise, it then uniformly selects $d_{k,t}$ symbols from its coded symbol vector, linearly combines them and then sends out the result over RE $t$. Denote the set of indices of the selected symbols as $\mathcal{V}(k,t)\subseteq\{1,2,...,N\}$.
Thus the signal sent by $U_k$ over RE $t$ is
 \begin{equation}\small
\begin{aligned} x'_{k,t}=
\begin{cases}
\sum_{j\in \mathcal{V}(k,t)}x_{k,j}&  d_{k,t}\not= 0\\
0& d_{k,t}= 0
\end{cases}.
\end{aligned}
\end{equation}
The signals transmitted over the same RE $t$ from all the users are linearly added together in the air, thus the signal received by the AP over RE $t$ can be written as
\begin{equation}\small
\begin{aligned}
\label{y_t}
   y_{t}=\sum_{k\in \{1,2,...K\}}h_{k}x'_{k,t}+z_t,
\end{aligned}
\end{equation}
where $h_k$ denotes the channel gain from $U_k$ to the AP, which is $0$ for inactive users, and $z_t$ denotes the Gaussian noise at RE $t$ with mean $0$ and variance $\xi_w$.

The AP consistently and coherently collects signals from the available REs and keeps attempting to detect the activity of the potential users and decode all the active users' messages until they are successfully retrieved or there are no active users, by viewing the transmission process as a special kind of \emph{linear superposition rateless encoder}. Once a user message is successfully decoded, the AP feeds back to that user an acknowledgement (ACK) to end its current transmission cycle.

\subsection{Joint active user detection and decoding algorithm}
\begin{figure}[!htp]
\centering
\includegraphics[width=0.45\textwidth]{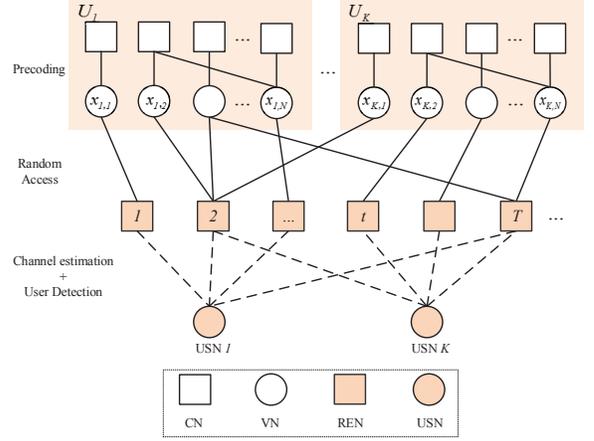}
\caption{Graph representation of the grant-free RMA. Each subgraph within a rectangle represents the LDPC code applied at a user. The edge between a VN and an REN means that the corresponding symbol is transmitted over that RE.
 Each USN connects to the set of RENs that the corresponding user pseudo-randomly selects during the access process.  The soft messages transmitted on solid and dashed edges are about information bits and the user status, respectively.}
\label{GFRMA}
\end{figure}
Following \cite{RMA}, the aforementioned process can be elegantly represented by a factor graph as depicted in Fig.\ref{GFRMA}. However, unlike \cite{RMA}, in which only three types of nodes, including check nodes (CNs), variable nodes (VNs) and resource element nodes (RENs), are involved, the user status of activity, indicated by its channel gain, is expressed as a special type of variable nodes called user status nodes (USNs). %, for that it is able to obtain information from them

The AP performs iterative detection and decoding on the graph. More specifically, by exchanging soft messages on the graph iteratively, the AP recovers VNs and USNs, which correspond to the information decoding, the channel estimation and active user detection respectively. %Utilizing the factor graph to realize the active user detection is novel in this paper.
%As the algorithm iterates, all the variable nodes
%In each iteration, VNs obtain soft message from the RENs, according to which, it tries to decode
Here we introduce some notations that will be used later. The superscript $n$ denotes the iteration $n$, $L_{(k,j)\xrightarrow[]{}t}^n$ denotes the log likelihood ratio passed from $j$th VN of $U_k$, namely VN $(k,j)$, to REN $t$, $(\mu_{h_{k}}^{n},\xi_{h_{k}}^{n})$ denotes the estimated mean and variance of the channel gain from $U_k$ to the AP, and $q_{k}^n$ denotes the estimated active probability of $U_k$ in the current transmission cycle.
In the initial stage, we have
\begin{align}\small
L_{(k,j)\xrightarrow[]{}t}^0=0,~(\mu_{h_{k}}^{0},\xi_{h_{k}}^{0})=(\bar{\mu}_{h_k},\bar{\xi}_{h_k}),~q_{k}^0=p_\text{a},
\end{align}
where $(\bar{\mu}_{h_k},\bar{\xi}_{h_k})$ is the initial mean and variance of User $k'$s
channel gain\footnote[1]{We assume that the AP knows some rough information about the mean
and variance of channel gains so as to start the iterative detection process. In
case they are unavailable, we can simply set them to be some proper non-zero
values (e.g. 1 and 10, respectively) and use the asymptotic LDPC codes to eliminate the phase ambiguity \cite{WO1}. The joint iterative algorithm can then
re-estimate their exact values.}, and $p_\text{a}$ is the average user active probability.

In the following, the iterative joint detection and decoding algorithm is described. First, we consider the messages passed from RENs to VNs. Let $\mathcal{M}(t)$ denote $\bigcup_{k\in\{1,...K\}}\mathcal{V}(k,t)$.  Thereby, (\ref{y_t}) can be equivalently reformulated as %In the decoding algorithm, RENs transmit soft messages to the connected USNs and VNs, which correspond to the CSI and the information
\begin{equation}\small
\begin{aligned}
\label{y_t_gau}
   y_{t}&= h_kx_{k,j}+\varsigma_{t|(k,j)}+z_t,
\end{aligned}
\end{equation}
where $\varsigma_{t|(k,j)}= \sum_{(k',j')\in\mathcal{M}(t)\backslash(k,j)}h_{k'}x_{k',j'}$. To reduce complexity, we resort to Gaussian Approximation (GA) as in \cite{IDMA}, and approximate $\varsigma_{t|(k,j)}$ to be Gaussian-distributed with mean $\mu_{\varsigma_{t|(k,j)}}$ and variance $\xi_{\varsigma_{t|(k,j)}}$ (please refer to \cite{IDMA} for details about them).
%Specifically, in iteration $n$, we have
%\begin{equation}\small	
%\begin{aligned}
%\label{ECS_mu}
%      \mu_{\varsigma_{t|(k,j)}}^n&= \sum_{(k',j')\in\mathcal{M}(t)\backslash(k,j)}\text{E}(h_{k'}x_{k',j'})\\
%      &=\sum_{(k',j')\in\mathcal{M}(t)\backslash(k,j)}{\mu_{h_{k'}}^nq_{k'}^n\text{tanh}\Big(\frac{L_{(k',j')\xrightarrow[]{}t}^n}{2}\Big)},
%\end{aligned}
%\end{equation}
%and
%\begin{equation}\small	
%\begin{aligned}
%\label{ECS_xi}
%       &\xi_{\varsigma_{t|(k,j)}}^{n}=\sum_{(k',j')\in\mathcal{M}(t)\backslash(k,j)}\text{E}\Big( h_{k'}x_{k',j'}-\text{E}(h_{k'}x_{k',j'})\Big)^2,\\
%       \end{aligned}
%\end{equation}
%respectively, where
%\begin{equation}\small	
%\begin{aligned}
%\label{ECS_xi2}
%&\text{E}\Big( h_{k'}x_{k',j'}-\text{E}(h_{k'}x_{k',j'})\Big)^2\\
% &= q_{k'}^n\Big[(\mu_{h_{k'}}^{n})^2+\xi_{h_{k'}}^{n}\Big]
%       -\Big(q_{k'}^n\mu_{h_{k'}}^{n}\text{tanh}(\frac{L_{({k'},j')\xrightarrow[]{}t}^n}{2})\Big)^2.
%       \end{aligned}
%\end{equation}
As such, (\ref{y_t_gau}) can be further formulated as
\begin{equation}\small		
\begin{aligned}
\label{y_t_new}
y_{t}=\mu_{h_k}^nx_{k,j}+z_{h_k}^n+\varsigma_{t|(k,j)}+z_t,
    \end{aligned}
\end{equation}
where $z_{h_k}^n\sim\mathcal{N}(0,\xi_{h_k}^n)$. Based on this, the message from REN $t$ to VN $(k,j)$ can be calculated by
\begin{equation}\small		
\begin{aligned}
\label{LLR}
    L_{t\xrightarrow[]{}(k,j)}^{n+1}=\log{\frac{p(y_t|x_{k,j}=1)}{p(y_t|x_{k,j}=-1)}}=\frac{2\mu_{h_k}^n(y_t-\mu_{\varsigma_{t|(k,j)}}^n)}{ \xi_{h_{k}}^{n}+ \xi_{\varsigma_{t|(k,j)}}^{n}+\xi_w}.
    \end{aligned}
\end{equation}

Second, the soft message that VN $(k,j)$ gets from all the connected RENs can be expressed as
\begin{equation}\small		
\begin{aligned}
\label{LLR_total}
 L_{k,j}^{n+1}=\sum_{t\in\mathcal{R}(k,j)} L_{t\xrightarrow[]{}(k,j)}^{n+1},
  \end{aligned}
\end{equation}
where $\mathcal{R}(k,j)$ denotes the set of indices of RENs that connect to VN $(k,j)$. Based on this, the LDPC coding part can be updated iteratively as in \cite{DE} and feeds back $L_{(k,j)\xrightarrow[]{}t}^{n+1}$ to aid the active user detection as described below.

Third, we approximate $h_k$ with soft messages received from RENs. From (\ref{y_t_gau}), employing similar approximation method as in \cite{WO}, we can get the symbol-wise channel estimate from messages passed from REN $t$ to USN $k$ as follows:
\begin{equation}\small		
\begin{aligned}
\label{u_h}
    \mu_{{h_k} \leftarrow t}^{n+1}=\frac{y_t-\mu_{\varsigma_{t|(k,j)}}^n}{\text{tanh}(\frac{L_{(k,j)\xrightarrow[]{}t}^n}{2})},
\end{aligned}
\end{equation}
and the estimation variance is
\begin{equation}\small		
\begin{aligned}
\label{sigma_h}
    \xi_{{h_k} \leftarrow t}^{n+1}=\frac{\xi_{\varsigma_{t|(k,j)}}^{n}+\xi_{w}}{\text{tanh}(\frac{L_{(k,j)\xrightarrow[]{}t}^n}{2})^2}.
\end{aligned}
\end{equation}
Denote the set of indices of RENs that connect to USN $k$ as $\mathcal{Q}(k)$. Using the result for the product of Gaussian distributions\cite{Gasu}, USN $k$ optimally combines all the messages from $\mathcal{Q}(k)$ as well as the initial channel information, and re-estimates the mean and variance of the channel gain as
%
%Using the result for product of Gaussian distributions, we have
%\begin{equation}		
%\begin{aligned}
%\label{in}
%(\mu_{h_{k,\text{RE}}}^{n+1},\xi_{h_{k,\text{RE}}}^{n+1})&=(\frac{\sum\limits_{t\in\mathcal{Q}(k)}\frac{\mu_{{h_k} \leftarrow t}^{n+1}}{\xi_{{h_k} \leftarrow t}^{n+1}}}{\sum\limits_{t\in\mathcal{Q}(k)}\frac{1}{\xi_{{h_k} \leftarrow t}^{n+1}}},\frac{1}{\sum\limits_{t\in\mathcal{Q}(k)}\frac{1}{\xi_{{h_k} \leftarrow t}^{n+1}}}).
%\end{aligned}
%\end{equation}
%Then we take the prior information into consideration, we can calculate $(\mu_{h_k}^{n+1},\xi_{h_k}^{n+1})$ as follows:
\begin{equation}\small		
\begin{aligned}
\label{ALL}
(\mu_{h_k}^{n+1},\xi_{h_k}^{n+1})&=(\frac{\sum\limits_{t\in{\mathcal{Q}}(k)}\frac{\mu_{{h_k} \leftarrow t}^{n+1}}{\xi_{{h_k} \leftarrow t}^{n+1}}+\frac{\bar{\mu}_{h_k}}{\bar{\xi}_{h_k}}}{\sum\limits_{t\in{\mathcal{Q}}(k)}\frac{1}{\xi_{{h_k} \leftarrow t}^{n+1}}+\frac{1}{\bar{\xi}_{h_k}}},\frac{1}{\sum\limits_{t\in{\mathcal{Q}}(k)}\frac{1}{\xi_{{h_k} \leftarrow t}^{n+1}}+\frac{1}{\bar{\xi}_{h_k}}}).
\end{aligned}
\end{equation}

Finally, based on the re-estimated mean and variance of the channel gain, USN $k$ calculates the active probability of $U_k$ using the MAP criterion:
%\begin{equation}\tiny		
%\begin{aligned}
%\label{ALLP}
%&q_k^{n+1}\\
%&=\frac{\frac{p_\text{a}}{\sqrt{\xi_{h_{k,\text{RE}}}^{n+1}+\bar{\xi}_{h_k}}}\exp\Big(-\frac{(\mu_{h_{k,\text{RE}}}^{n+1}-\bar{\mu}_{h_k})^2}{2(\xi_{h_{k,\text{RE}}}^{n+1}+\bar{\xi}_{h_k})}\Big)}{\frac{p_\text{a}}{\sqrt{\xi_{h_{k,\text{RE}}}^{n+1}+\bar{\xi}_{h_k}}}\exp\Big(-\frac{(\mu_{h_{k,\text{RE}}}^{n+1}-\bar{\mu}_{h_k})^2}{2(\xi_{h_{k,\text{RE}}}^{n+1}+\bar{\xi}_{h_k})}\Big)+\frac{1-p_\text{a}}{\sqrt{\xi_{h_{k,\text{RE}}}^{n+1}}}\exp\Big(-\frac{
%(\mu_{h_{k,\text{RE}}}^{n+1})^2}{2\xi_{h_{k,\text{RE}}}^{n+1}}\Big)}.
%\end{aligned}
%\end{equation}
\begin{equation}\small		
\begin{aligned}
\label{ALLP}
q_k^{n+1}=\frac{\frac{p_\text{a}}{\sqrt{\bar{\xi}_{h_k}}}\exp\Big(-\frac{(\mu_{h_{k}}^{n+1}-\bar{\mu}_{h_k})^2}{2\bar{\xi}_{h_k}}\Big)}{\frac{p_\text{a}}{\sqrt{\bar{\xi}_{h_k}}}\exp\Big(-\frac{(\mu_{h_{k}}^{n+1}-\bar{\mu}_{h_k})^2}{2\bar{\xi}_{h_k}}\Big)+\frac{1-p_\text{a}}{\sqrt{\xi_{h_{k}}^{n+1}}}\exp\Big(-\frac{
(\mu_{h_{k}}^{n+1})^2}{2\xi_{h_{k}}^{n+1}}\Big)}.
\end{aligned}
\end{equation}
Note that (\ref{ALL}) and (\ref{ALLP}) are used in the iterative decoding part such as in (\ref{LLR}) and the calculation of $\mu_{\varsigma_{t|(k,j)}}$ and $\xi_{\varsigma_{t|(k,j)}}$.

%In practice, the detection of active users usually takes only the first a few iterations, and the AP need not to update the factor graph of the inactive users, thus the complexity can be greatly reduced.

\section{Performance Analysis} \label{analysis}
The average system throughput $\mathcal{T}=\frac{Kp_\text{a}m}{T}$ and the average signal-to-noise ratio (SNR) over all the REs, denoted by $\gamma$, can be calculated by $\gamma=\frac{\sum_{k\in\{1,2,...K\}}\text{E}(d_{k,t})h_k^2}{\xi_w}$. The main object of this section is to derive the relationship between the system throughput and the corresponding threshold SNR, above which the error rate is able to converge to zero as the algorithm iterates. The DE analysis is an effective approach to analyze the performance of iterative decoding algorithms in the asymptotic sense, i.e., the quantities of each type of nodes in the decoding graph tend to infinity \cite{DE}. However, in our model, as depicted in Fig.\ref{GFRMA}, each user only has one USN, which does not satisfy the asymptotic condition, and the channel estimate distribution cannot be used as in the conventional DE.

More specifically, to estimate the decoding performance based on the DE analysis\cite{DE}, we need to analyze the evolution process of the mutual information (MI)%between the soft messages that VNs get from the connected RENs, $L_{k,j}^{n+1}$, and the corresponding variable, $x_{k,j}$, during the iterations. For clarity, the MI is denoted by
\begin{equation}\small		
\begin{aligned}
\label{MI_EV}
I(L_{k,j}^{n+1};x_{k,j}).
\end{aligned}
\end{equation} %Denote $\mu_{h_{k,la}}^n$ as the channel estimation that is \emph{least appropriate} for the decoding.
According to (\ref{LLR})-(\ref{LLR_total}), the relationship between $L_{k,j}^{n+1}$ and $x_{k,j}$ can be expressed as:
\begin{equation}\small		
\begin{aligned}
\label{MI}
 L_{k,j}^{n+1}&=\sum_{t\in\mathcal{R}(k,j)}\Big({\frac{2\mu_{h_{k}}^nh_kx_{k,j}+2\mu_{h_{k}}^n(\varsigma_{t|(k,j)}-\mu_{\varsigma_{t|(k,j)}}+z_t)}{\xi_{h_k}^n+\xi_{\varsigma_{t|(k,j)}}^n+\xi_w}}\Big)\\
             %&\quad\quad\quad\quad\quad\quad +{\frac{2\mu_{h_{k}}^n(\varsigma_{t|(k,j)}-\mu_{\varsigma_{t|(k,j)}}+z_t)}{\xi_{h_k}^n+\xi_{\varsigma_{t|(k,j)}}^n+\xi_w}}\Big)\\
             &\triangleq \mu_{L_{k}}^{n+1}x_{k,j}+z_{L_{k}}^{n+1},&
\end{aligned}
\end{equation}
where $z_{L_{k}}^{n+1}$ is Gaussian distributed with mean $0$ and variance $\xi_{L_{k}}^{n+1}$. Note that $L_{k,j}^{n+1}$ is also influenced by the soft message of the channel gain, i.e., $(\mu_{h_{k}}^{n},\xi_{h_{k}}^{n})$. Since $\xi_{h_k}^n\neq 0$ due to the relatively short block length, the distribution of $L_{k,j}^{n+1}$ %for the specific $\mu_{h_{k}}^n$
is not symmetric\cite{MIS2}, i.e., $\mu_{L_{k}}^{n+1}\neq \frac{\xi_{L_{k}}^{n+1}}{2}$. Furthermore, the estimated mean of the channel gain, $\mu_{h_{k}}^n$, is a \emph{random variable}, and the MI
defined by (\ref{MI_EV}) in each iteration is also a random variable, which makes the direct application of the conventional DE impossible. For rescue, we propose to use $\text{E}_{\mu_{h_{k}}^n}(I(L_{k,j}^{n+1};x_{k,j}))$ instead.
%In the following, we propose an approximate method to calculate it. %based on which, the threshold SNR can be estimated.
%More specifically, in the first step, we approximates the distribution of $\mu_{h_{k}}^n$. In the second, we derive the MI between $L_{k,j}^{n}$, and $x_{k,j}$ for the given $\mu_{h_{k}}^n$, i.e., $I(L_{k,j}^{n+1};x_{k,j}|\mu_{h_{k}}^n)$.

To this end, we first need to obtain the distribution of $\mu_{h_{k}}^n$. %according to (\ref{MI}) and the above discussions, the MI defined by (\ref{MI1}) is also a random variable. %which makes the direct application of the conventional DE impossible.
Since it is generally hard to get the exact distribution of $\mu_{h_k}^n$ based on the above iterative process, by using the fact that its distribution well approximates the Gaussian distribution and tends to $\mathcal{N}(h_k,\xi_{h_k}^n)$ when the MI of VNs approaches $1$, we assume that $\mu_{h_k}^n\sim\mathcal{N}(h_k,\xi_{h_k}^n)$. Then, we derive $I(L_{k,j}^{n+1};x_{k,j}|\mu_{h_{k}}^n)$. %which denotes the MI between $L_{k,j}^{n+1}$ and $x_{k,j}$ with specific $\mu_{h_k}^n$ and
More specifically, we consider it under two cases.
First, in the case in which $\mu_{h_k}^n\leq h_k$, based on (\ref{MI}), we have $\mu_{L_k}^{n+1}>\frac{\xi_{L_k}^{n+1}}{2}$, and thus $L_{k,j}^{n+1}$ supplies more information to VNs than the message with distribution $\mathcal{N}(\mu_{L_{k}}^{n+1}x_{k,j},2\mu_{L_{k}}^{n+1})$.
On the other hand, in the case $\mu_{h_k}^n>h_k$, since the BP decoding of LDPC codes is less sensitive to the overestimation than it is to the underestimation of channel gains\cite{MIS2}, the decoding performance is better than that in the contrary case. %Thereby, in all cases, $I(L_{k,j}^{n+1};x_{k,j}|\mu_{h_{k}}^n)$ can be lower bounded by the MI of a symmetric Gaussian distribution.

At last, since $\xi_{\varsigma_{t|(k,j)}}^n$ defined in (\ref{MI}) is expected to be independent of $t$ and $(k,j)$, due to the law of large numbers, we simply define $\xi_{\varsigma}^n\triangleq\xi_{\varsigma_{t|(k,j)}}^n$. Thus, we have
\begin{equation}\small
\begin{aligned}
\mu_{L_{k}}^{n+1}=\frac{2h_k\text{E}(d_{k,t})T}{\Big(\xi_{\varsigma}^n+\xi_{h_k}^n+\xi_w\Big)N}\mu_{h_{k}}^{n}\triangleq \mathcal{L}_1(\mu_{h_{k}}^{n})
\end{aligned}.
\end{equation}
Since the decoding performance of LDPC codes in the case $\mu_{h_k}^n\leq h_k$ is worse than that of the contrary case\cite{MIS2}, we only consider the case $\mu_{h_k}^{n}\leq h_k$ approximately. Thus, we have
\begin{equation}\small
\begin{aligned}
\text{E}_{\mu_{h_{k}}^n}(I(L_{k,j}^{n+1};x_{k,j}))\ge\int_{-\infty}^{h_k}I(L_{k,j}^{n+1};x_{k,j}|\mu_{h_k}^n)\cdot2p(\mu_{h_k}^n){\rm d}\mu_{h_k}^n.
\end{aligned}
\end{equation}
Since $I(L_{k,j}^{n+1};x_{k,j}|\mu_{h_{k}}^n)$ in this case can be lower bounded by the MI of a symmetric Gaussian distribution, we approximately model $L_{k,j}^{n+1}$ by a random variable that is distributed as $\mathcal{N}(\mu_{L_{k}}^{n+1}x_{k,j},2\mu_{L_{k}}^{n+1})$, and by doing so the analysis of LDPC coding part can be done with reduced complexity. Thereby,
\begin{equation}\small
\begin{aligned}
\text{E}_{\mu_{h_{k}}^n}(I(L_{k,j}^{n+1};x_{k,j}))\ge\int_{-\infty}^{h_k}J\Big(\sqrt{2\mathcal{L}_1(\mu_{h_{k}}^{n})}\Big)\cdot2p(\mu_{h_k}^n){\rm d}\mu_{h_k}^n,
\end{aligned}
\end{equation}
where $J(x)$ is defined as
\begin{equation}\small
\begin{aligned}
  &J(x) =1-\int_{-\infty }^{+\infty}\frac{1}{\sqrt{2\pi x^2}}\exp{\{-\frac{(\xi-\frac{x^2}{2})^2}{2x^2}\}}\cdot\log_2(1+e^{-\xi}){\rm d}\xi. \\
\end{aligned}
\end{equation}
Furthermore, let $d_v$ denote the average degree of VNs of the adopted LDPC codes and $\mu_{L_{{\text{C}\xrightarrow[]{}\text{V}}_k}}^{n+1}$ denote the mean of the log likelihood ratio passed from the CNs to VNs of $U_k$, which can be calculated based on the aforementioned $\mu_{L_{k}}^{n+1}$ and $\mu_{h_k}^n$ (please be referred to \cite{DE} for details). For clarity, we define
the calculation of $\mu_{L_{{\text{C}\xrightarrow[]{}\text{V}}_k}}^{n+1}$ as $\mathcal{L}_2(\mu_{h_k}^n)$. As a result, the MI of VNs of  $U_k$ in iteration $n+1$ can be expressed as follows:
\begin{equation}\small		
\begin{aligned}
\label{j_func}
I_k^{n+1}\triangleq\int_{-\infty}^{h_k}J\Big(\sqrt{2\Big(\mathcal{L}_1(\mu_{h_{k}}^{n})+d_v\mathcal{L}_2(\mu_{h_k}^n)\Big)}\Big)\cdot2p(\mu_{h_k}^n){\rm d}\mu_{h_k}^n
.
\end{aligned}
\end{equation}
%Furthermore, based on the DE analysis\cite{DE}, if the MI of VNs converges to $1$, the error rate converges to zero gradually as the algorithm iterates \cite{DE}.
%However, since the estimated mean of the channel gain, $\mu_{h_{k}}^n$, is a \emph{random variable} due to the relatively short block length, according to (\ref{MI}) and the above discussions, the MI defined by (\ref{j_func}) in each iteration is also a random variable, which makes the direct application of the conventional DE impossible. For rescue, we propose to use the estimated mean that supplies the \emph{least information} to the LDPC decoder, which is denoted as $\mu_{{h_k}|\textrm{LI}}^n$ and has a fixed value for each iteration $n$, to replace the original $\mu_{h_{k}}^n$ in (\ref{MI}). In the following, we denote the corresponding $\mu_{L_{k}}^{n+1}$ as $\mu_{L_{k}|\textrm{LI}}^{n+1}$.
%Now the DE analysis can be used to analyze the MI of VNs, and as a result the upper bound of threshold SNR or the worst-case threshold can be obtained.
As such, %based on the DE\cite{DE},
the threshold SNR $\gamma_{th}$ can be expressed as
\begin{equation}\small		
\begin{aligned}
\gamma_{th}=\min\{\gamma: I_k^\infty=1,k\in \mathcal{K}_a\},
\end{aligned}
\end{equation}
where $\mathcal{K}_a$ is the set of indices of active users.

\section{Simulation Results} \label{numer}
In the simulation, we set that there are $100$ potential users. Among them, $10$ randomly chosen ones are active (i.e., $p_\text{a}=0.1$) and each transmits a packet of $240$ bits to the AP. We set the total number of REs as $6400$ and consider the scenario where the channels from the active users to the AP are the same. The initial mean and variance of the channel gains, $(\bar{\mu}_{h_k},\bar{\xi}_{h_k})$, are all set as $(1,10)$. Following the optimization in \cite{RMA}, we choose RACf $\rho(x)=0.06x+0.04$ and that each active user encodes its packet with an LDPC code of rate $0.6$. % and thus the average throughput is $0.375$ bits per RE.
\begin{figure}[!ht]
\centering
\includegraphics[width=3.2in]{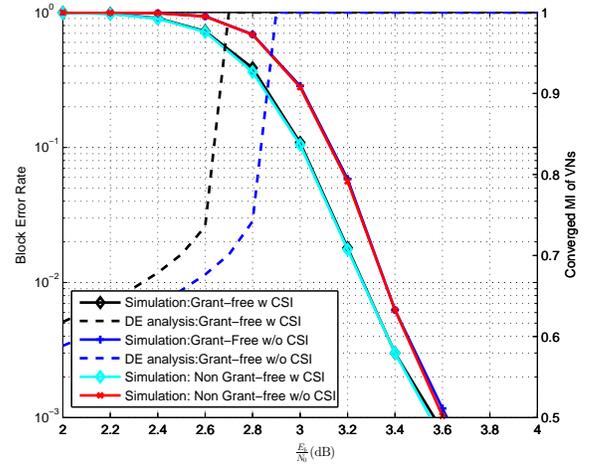}
\caption{The converged MI (dashed lines) and BLER of the proposed grant-free RMA: compared with the registration-based RMA, the proposed scheme requires no registration and has only a little sacrifice in BLER performance.}
\label{Simulation_GFRMA}
\end{figure}
%The dashed lines in Fig.\ref{Simulation_GFRMA} represent the analytical results of the converged MI for the given throughput and $\frac{E_b}{N_0}$, based on which, the threshold for the two cases are $2.7$dB and $2.9$dB. %The analytical results for grant-free RMA without CSI is a little looser since the symmetrical distribution is taken as the lower bound.
As illustrated in Fig.\ref{Simulation_GFRMA}, analytical and simulation results are consistent in demonstrating that the grant-free RMA only has a very little sacrifice in the block error rate (BLER) performance.

\section{Conclusion} \label{conclusion}
In this work, a grant-free RMA scheme is proposed to address the challenge brought by short-packet and low-latency transmissions in IoT networks. Our scheme omits the requirement of registration with a very little performance degradation, thus is a viable candidate for massive access in the IoT system. Furthermore, we propose a method based on DE to theoretically analyze the system performance, which gives instructions to the practical system design. At last, extending the proposed framework to the case with fast fading channels is left as a future work \cite{LMMSE}.

\bibliographystyle{IEEEtran}

\begin{thebibliography}{1}
\bibitem{M2M}
A. Zanella, M. Zorzi, A. F. dos Santos, P. Popovski, etc., ``M2M massive
wireless access: Challenges, research issues, and ways forward,'' in \emph{ Proc. IEEE Globecom Workshops},  2013.

\bibitem{Short}
G. Durisi,  T. Koch, and  P. Popovski, ``Towards Massive, Ultra-Reliable, and Low-Latency Wireless: The Art of Sending Short Packets,'' arXiv:1504.06526.

\bibitem{LTE}
E. Dahlman, S. Parkvall, and J. Skold, ``4G: LTE/LTE- Advanced for Mobile Broadband,'' 2nd ed. Waltham, MA, USA: Elsevier, 2014.


\bibitem {RMA}
X. Wang, Z. Zhang, Y. Zhang, L. Zhang, and Y. Chen, ``Multi-Carrier Rateless Multiple Access: A Novel Protocol for Dynamic Massive Access,'' in \emph{ Proc. IEEE Globecom}, 2015.

\bibitem{Zhu}
H. Zhu,  and  G. B. Giannakis,  ``Exploiting Sparse User Activity in Multiuser Detection,''  \emph{IEEE Trans. Commun.},  2011.

\bibitem{Bockelmann1}
F. Monsees, C. Bockelmann,  D. Wubben, and A. Dekorsy, ``Sparsity Aware Multiuser detection for Machine to Machine communication,'' in \emph{Proc. IEEE Globecom Workshops},  2012.


\bibitem{Bockelmann3}
C. Bockelmann, ``Iterative Soft Interference Cancellation for Sparse BPSK Signals,'' \emph{IEEE Commun. Lett.}, 2015.

\bibitem {WO1}
 T. Wo, C. Liu, and P. A. Hoeher, ``Graph-Based Soft Channel and Data Estimation for MIMO Systems with Asymmetric LDPC Codes,''  in \emph{Proc. IEEE ICC}, 2008.

%\bibitem{Neighbor}
%L. Zhang, J. Luo, and D. Guo, ``Neighbor discovery for wireless networks via compressed sensing,'' \emph{Performance Evaluation}, 2013.

%\bibitem {Sparse}
%G. Wunder, H. Boche, T. Strohmer, and P. Jung, ``Sparse Signal Processing Concepts for Efficient 5G System Design,'' arXiv:1411.0435.

\bibitem{IDMA}
L. Ping, L. Liu, K. Wu, and W. K. Leung, ``Interleave division multiple access,'' \emph{IEEE Trans. Wireless Commun.}, 2006.

\bibitem{DE}
J. Chen, and M. Fossorier, ``Density evolution for two improved BP-based decoding algorithms of LDPC codes,'' \emph{IEEE Commun. Lett.}, 2002.

\bibitem {WO}
E. Aktas, ``Iterative Message Passing for Pilot-Assisted Multiuser Detection in MC-CDMA Systems,'' \emph{IEEE Trans. Commun.}, 2012.

\bibitem {Gasu}
J. Pearl, ``Probabilistic Reasoning in Intelligent Systems: Networks of Plausible Inference.'' San Francisco, CA: Morgan Kaufmann Publishers, 1988.

\bibitem{MIS2}
H. Saeedi, and A. H. Banihashemi, ``Performance of Belief Propagation for Decoding LDPC Codes in the Presence of Channel Estimation Error,'' \emph{IEEE Trans. Commun.}, 2007.

\bibitem{LMMSE}
K. Takeuchi, R. R. M\"uller, and M. Vehkaper\"a, ``Iterative LMMSE Channel Estimation and Decoding Based on Probabilistic Bias,'' \emph{IEEE Trans. Commun.}, 2013.
\end{thebibliography}

\end{document}